\documentclass{svproc}
\usepackage{url}

\usepackage{graphicx}
\begin{document}
\mainmatter              
\title{Complex Network Analysis of North American Institutions of Higher Education on Twitter}
\titlerunning{Network of Colleges on Twitter}  
%
\author{Dmitry Zinoviev \and Shana Cote \and Robert D\'iaz}
\authorrunning{Zinoviev et al.} 
%
\tocauthor{Dmitry Zinoviev, Shana Cote, and Robert D\'iaz}
\institute{Suffolk University, Boston MA 02114, USA,\\
\email{dzinoviev@suffolk.edu}, \email{\{scote4,rdiaz2\}@su.suffolk.edu}}

\maketitle              

\begin{abstract}
North American institutions of higher education (IHEs): universities,
4- and 2-year colleges, and trade schools---are heavily present and
followed on Twitter. An IHE Twitter account, on average, has 20,000
subscribers. Many of them follow more than one IHE, making it possible
to construct an IHE network, based on the number of co-followers. In
this paper, we explore the structure of a network of 1,435 IHEs on
Twitter. We discovered significant correlations between the network
attributes: various centralities and clustering coefficients---and
IHEs' attributes, such as enrollment, tuition, and
religious/racial/gender affiliations. We uncovered the community
structure of the network linked to homophily---such that similar
followers follow similar colleges. Additionally, we analyzed the
followers' self-descriptions and identified twelve overlapping topics
that can be traced to the followers' group identities.
\end{abstract}
\section{Introduction}
According to the National Center for Education
Statistics~\cite{nces2018}, in 2018, there were 4,313 degree-granting
postsecondary institutions, also known as institutions of higher
education (IHEs), in the USA. This number includes public and private
(both nonprofit and for-profit) universities, liberal arts colleges,
community colleges, religious schools, and trade schools.

The IHEs enjoy a heavy presence on social media, in particular, on
Twitter.  In 2012, Linvill {\em et al.}~\cite{linvill2012} found that
IHEs employ Twitter primarily as an institutional news feed to a
general audience. These results were confirmed by Kimmons {\em et al.}
in 2016~\cite{kimmons2016} and 2017~\cite{veletsianos2017}; the
authors further argue that Twitter failed to become a ``vehicle for
institutions to extend their reach and further demonstrate their value
to society''---and a somewhat ''missed opportunity for presidents to
use Twitter to connect more closely with alumni and
donors''~\cite{walton2020}. The same disconnect has been observed for
IHE library accounts~\cite{stewart2018}.

Despite the failed promise, the IHEs massively invest in online
marketing~\cite{taylor2020} and, in reciprocity, collect impressive
follower lists that include both organizations and individuals. The
longer follower lists demonstrate a positive effect on IHE
performance, particularly, on student recruitment~\cite{rutter2016},
and may eventually affect IHE ratings or at least correlate with
them~\cite{mccoy2017}. Therefore, follower lists are essential
marketing instruments and should be studied comprehensively.

To the best of our knowledge, this paper is the first attempt to look
at a social network of IHE Twitter accounts based on the similarities
of their follower lists. We hypothesize that the exogenous
parameters, such as enrollment, tuition, and religious/gender/race
preferences, affect the structure of the network and
positions/importance of the IHEs in it.

The rest of the paper is organized as follows: In
Section~\ref{sec:data}, we describe the data set, its provenance, and
structure; in Section~\ref{sec:construction}, we explain the network
construction; in Section~\ref{sec:analysis}, we go over the network
analysis, and present the results; in Section~\ref{sec:glue}, we take
a look at the followers; in Section~\ref{sec:discussion}, we discuss
the results.  Finally, in Section~\ref{sec:conclusion}, we conclude.

\section{\label{sec:data}Data Set}
Our data set consists of two subsets: social networking data from
Twitter and IHE demographics from Niche~\cite{niche_com}. We used the
former to construct a network of IHEs and the latter to provide
independent variables for the network analysis. Both subsets were
collected in Summer 2020.

The Twitter data set describes the Twitter accounts of 1,450 IHEs from
all 50 states and the District of Columbia. The majority of the
accounts are the official IHE accounts, but for some IHEs, we had to
rely on secondary accounts, such as those of admission offices or
varsity sports teams. For each IHE, we have the following attributes
(and their mean values): geographical location (including the state),
the lists of followers (20,198) and friends (1,130), the numbers of
favorites (``likes''; 4,656) and statuses (``posts''; 9,132), the
account age in years (10.4), and whether the account is verified or
not (32\% accounts are verified).

With some IHEs having more than a million followers (e.g., MIT and
Harvard University), we chose to restrict our lists to up to 10,000
followers per IHE. This limitation may have resulted in a slight
underestimation of the connectedness of the most popular IHEs. We
explain in subsection~\ref{subsection:weights} why we believe that the
underestimation is not crucial.

It is worth noting that while we have downloaded the friend lists, we
do not use them in this work because they are controlled by the IHE
administrations/PR offices and cannot be considered truly exogenous.


The combined list of followers consists of 347,920 users. This number
does not include the ``occasional'' followers who subscribed to fewer
than three IHEs.

The descriptive IHE data comes from Niche~\cite{niche_com}, an
American company that provides demographics, rankings, report cards,
and colleges' reviews. It covers 1,435 of the IHEs that we selected
for the network construction. Five more IHEs were not found on Niche
and, though included in the network, were not used in further
analysis.

For each IHE, we have the following attributes:

\begin{description}
\item[Binary:]
  \begin{itemize}
  \item ``Liberal Arts'' college designation,
  \item Application options: ``SAT/ACT Optional,'' ``Common App
    Accepted,'' or ``No App Fee'' (these options can be combined).
  \end{itemize}
\item[Categorical:]
  \begin{itemize}
  \item Type: ``Private,'' ``Public,'' ``Community College,'' or
    ``Trade School''; note that all community colleges and trade
    schools in our data set are public;
  \item Religious affiliation: ``Christian,'' ``Catholic,'' ``Muslim''
    or ``Jewish''; we lumped the former two together;
  \item Online learning options: ``Fully Online,'' ``Large Online Program,''
    or ``Some Online Degrees'';
  \item Gender preferences: ``All-Women'' or ``All-Men'';
  \item Race preferences: ``Hispanic-Serving Institution'' (HSI) or
    ``Historically Black College or University'' (HBCU).
  \end{itemize}
\item[Count or real-valued:] Enrollment and tuition. We noticed that
  due to the broad range of enrollments and tuition%
, enrollment and tuition logarithms are
  better predictors. We will use
  $\log{\left(\mathrm{enrollment}\right)}$ and
  $\log{\left(\mathrm{tuition}\right)}$ instead of enrollment and
  tuition throughout the paper.
\end{description}


\section{\label{sec:construction}Network Construction}
We define the network $G$ of IHEs on
Twitter as $G=\left(N,E\right)$. Here, $N=\{n_i\}$ is a set of 1,450
nodes, each representing an IHE account, and $E=\{e_{ij}\}$ is a set
of weighted edges.

Let $\mathrm{f}(n)$ be a set of followers of the account $n$. As noted
in Section~\ref{sec:data}, $\forall n\in N: \#\mathrm{f}(n)\le10,000$.

Let $\mathrm{f}^{-1}(q)=\{n\in N\,|\,q\in\mathrm{f}(n)\}$ be a set of
all IHE accounts followed by user $q$. Note that $q$ itself may be a
member of $N$: IHEs can follow each other.

The definition of an edge is derived from the concept of $G$ as a
network based on co-following: two nodes $n_i$ and $n_j$ share an edge
$e_{ij}$ iff they have at least one shared follower that also follows
at least three IHE accounts. We denote a set of such qualified
followers as $Q$:
\begin{eqnarray}
  Q=\{q\,|\,\#\mathrm{f}^{-1}\left(q\right)\ge3\}\\
  \forall i,j:\,\exists e_{ij} \Leftrightarrow
  Q\cap\mathrm{f}(n_i)\cap\mathrm{f}(n_j)\ne\emptyset
\end{eqnarray}

The number of edges in $G$ is, therefore, 928,476. The network is
connected (there is only one connected component) and quite dense:
its density is 0.88.

Finally, let $w_{ij}>0$ be the weight of the edge $e_{ij}$. We
initially define $w_{ij}$ as the number of qualified shared followers:

\begin{equation}
w_{ij} = \#\left(Q\cap\mathrm{f}(n_i)\cap\mathrm{f}(n_j)\right).
\end{equation}


The resulting weights are large (on the order of $10^3$--$10^4$),
while many network algorithms, such as community detection and
visualization, expect them to be in the range $(0\ldots1]$. We used
  the algorithm proposed in~\cite{simas_rocha_2015} to normalize the
  weights without affecting the calculated node attributes.
  
\subsection{\label{subsection:weights}A Note on Edge Weight Calculations}
We mentioned in section~\ref{sec:data} that we use only up to 10,000
followers for edge weight calculations. The truncated follower lists
result in lower weights. We can estimate the difference between true
and calculated weights by assuming that the shared followers are
uniformly distributed in the follower lists. Let
$F=\overline{\#f}=21,123$ be the mean number of followers; let
$T=10,000$; let $p\approx0.685$ be the probability that a follower
list is not longer than $T$; let $\overline{w}$ be the mean edge
weight; finally, let $\overline{w^*}$ be the estimated mean edge
weight. Note that if $p=1$ then $\overline{w^*}=\overline{w}$. One can
show that:
\begin{equation}
  \frac{\overline{w^*}}{\overline{w}} \approx
  \left(\frac{\left(F-T\right)p+T}F\right)^2\approx
  1.436.
\end{equation}

Seemingly, the weights of all edges that are incident to at least one
node with a truncated follower list are underestimated by
$\approx30\%$. However, we noticed that Twitter reports follower lists
not uniformly but roughly in the order of prominence: the prominent
followers with many followers of their own are reported first. We hope
that the shared users responsible for edge formations are mostly
reported among the first 10,000 followers.

\section{\label{sec:analysis}Network Analysis}
In this section, we analyze the constructed network and present the
results. We looked at individual nodes' positions in the network
(monadic analysis), relations between adjacent nodes (dyadic
analysis), and node clusters (community analysis).

\subsection{\label{subsec:monadic}Monadic Analysis}
We used Python library {\em networkx}~\cite{zinoviev2018} to calculate
the monadic attributes: degree, closeness, betweenness, and
eigenvector centralities, and local clustering coefficient---for each
node $n\in G$. All the centralities of $n$ express various aspects of
$n$'s prominence in a network~\cite{wasserman1994}: the number of
closely similar IHEs (degree), the average similarity of $n$ to all
other IHEs (closeness), the number of IHEs that are similar to each
other by being similar to $n$ (betweenness), and the measure of mutual
importance (eigenvector: ``$n$ is important if it is similar to other
important nodes). The local clustering coefficient reports if the
nodes similar to $n$ are also similar to each other.



We use multiple ordinary least squares (OLS) regression to model the
relationships between each of the network attributes and the following
independent variables: tuition, enrollment, Twitter account age,
Twitter account verified status, ``No App Fee,'' ``Liberal Arts''
designation, ``SAT/ACT Optional,'' ``Common App Accepted,'' race
preferences, online learning options, type/religious affiliations, and
gender preferences (see Section~\ref{sec:data}). We combined the IHE
type and religious affiliations into one variable because all public
schools are secular.

The number of samples in the regression is 1,348 (the intersection of
the Niche set and Twitter set). Table~\ref{table:monadic} shows the
independent variables that significantly ($p\le0.01$) explain the
monadic network measures, and the regression coefficients.

\begin{table}
\caption{\label{table:monadic}Variables that significantly
  ($p\le0.01$) explain the monadic network measures: betw[enness],
  clos[eness], degr[ee], eigen[vector] centralities, clust[ering]
  coefficient, and numbers of favorites (``likes''), followers,
  friends, and statuses (posts). $^\dagger$The marked rows represent
  levels of the categorical variables.}
\begin{center}
\begin{tabular}{lrrrrrrrrrr}\hline
  Variable &  \multicolumn{9}{c}{Coef.}\\[2pt]
  \hline
 & betw. & clos. & clust. & degr. & eigen. & favorites & followers & friends & posts\\[2pt]\hline
Liberal Arts &  0.27 & 0.05 &  & 0.08 & 0.08 &  & -0.98 &  &  \\
Private$^\dagger$ & 0.29 &  &  &  &  &  & 1.40 &  &  \\
Account Age & 0.12 & 0.02 &  & 0.03 & 0.03 &  &  &    & 0.05\\
Tuition & -0.21 &  &  &  &  &  &  &  &   \\
Common App &  & 0.02 &  &  &  &  &  &  &   \\
No App Fee &  & 0.03 &  & 0.05 & 0.05 &  &  &  &  \\
Large Online$^\dagger$ &  & 0.05 &  & 0.09 & 0.09 &  &  &  & \\ 
Some Online$^\dagger$ &  & 0.02 &  & 0.04 & 0.04 &  & -0.55 &  & \\
HBCU$^\dagger$ &  & 0.06 &  & 0.11 & 0.10 &  &  &  &  \\
Christian$^\dagger$    & -0.04 & & 0.07 & 0.07 &  &  &   &  &  0.63  \\
Verified &  & -0.03 &  & -0.05 & -0.05 & 0.66 & 1.30 & 0.52  & 0.48\\
Enrollment &  & 0.03 & 0.01 & 0.06 & 0.06 & 0.33 & 0.65 & 0.29 & 0.29\\[2pt]
\hline
\end{tabular}
\end{center}
\end{table}

\subsection{Dyadic Analysis}
The only dyadic variable in our model is the edge weight. As a
reminder, the weight of an edge is derived from the number of Twitter
co-followers of the incident nodes. A stronger edge indicates a larger
overlap of the followers and, presumably, a closer similarity between
the IHEs, even if the nature and reason for the similarity is unclear.

We hypothesize that, because of homophily, edge weights depend on the
difference between the incident node attributes. We calculate the
dyadic versions of the monadic independent variables for the OLS
regression modeling as follows:

\begin{description}
\item[For the binary and categorical variables:] A calculated dyadic
  variable $y$ equals 1 if the values of the underlying monadic
  variable $x$ differ, and 0, otherwise:

  \begin{equation}
    y_{ij} = \left\{
    \begin{array}{ll}
      0 & \mbox{if } x_i=x_j \\
      1 & \mbox{if } x_i\ne x_j
    \end{array}
    \right.
  \end{equation}

  For example, if both incident nodes represent liberal art colleges,
  then the dyadic ``Same Liberal Arts designation'' variable for the
  edge is 0.
\item[For the count or real-valued variables:] A calculated dyadic
  variable $y$ equals the absolute value of the arithmetic difference
  of the underlying monadic $x$ variable at the incident nodes:
  \begin{equation}
    y_{ij} = |x_i-x_j|.
  \end{equation}
\end{description}

Both clauses emphasize the difference of the monadic attributes along
the incident edge. Table~\ref{table:edges} shows the independent
variables that significantly ($p\le0.01$) explain the edge weights, and
the regression coefficients. For this analysis, we add the state in
which an IHE is located to the monadic variables listed in
subsection~\ref{subsec:monadic}.

\begin{table}
\caption{\label{table:edges}Variables that significantly ($p\le0.01$)
  explain the edge weights}
\begin{center}
\begin{tabular}{lr}\hline
  Variable & Coef.\\[2pt]
  \hline
  Same state & 0.0169\\
  Similar enrollment & 0.0024\\
  Similar tuition  & 0.0022\\
  Same religious affiliation & 0.0019\\
  Same online preferences & 0.0010\\
  Similar account age  & 0.0008\\
  Same ``Common App Accepted'' option  & 0.0008\\
  Same ``No App Fee'' option  & 0.0008\\
  Same race designation  & 0.0006\\
  Same ``SAT/ACT Optional'' option  & 0.0005\\
  \hline
  Both verified  & -0.0001\\
  Same gender designation & -0.0022\\
  Same ``Liberal Arts'' designation  & -0.0026\\[2pt]
  \hline
\end{tabular}
\end{center}
\end{table}

\subsection{Community Analysis}

We used the Louvain community detection algorithm~\cite{blondel08} to
partition $G$ into network communities, or clusters: tightly connected
non-overlapping groups of nodes with more internal connections than
external connections. We requested a resolution of 0.8 (lower than the
standard 1.0) to discover smaller clusters and, as a result,
partitioned $G$ into 22 disjoint clusters $C=\{c_i\}$. The Newmann
modularity~\cite{newman2006} of the partition is 0.152 on the scale
$[-1/2\ldots1]$. Each cluster contains the nodes representing the
IHEs that are somewhat more similar to each other than to an IHE from
another cluster. In other words, the level of homophily within a
cluster is higher than between the clusters. We expect to identify the
independent variables responsible for the homophily. 

Table~\ref{table:cluster} shows the independent variables that
significantly ($p\le0.01$) explain the membership in select clusters,
and the regression coefficients. Note that the clusters 6, 9, 10, 16,
and 19 do not have any significant explanatory variables, and the
clusters 18, 20, 21, and 22 are single-node isolates.

\begin{table}
\caption{\label{table:cluster}Variables that significantly
  ($p\le0.01$) explain membership in select clusters. (See
  Fig.~\ref{figure:induced}.) $^\dagger$The marked rows represent
  levels of the categorical variables.}
\begin{center}
\begin{tabular}{lrrrrrrrrrrrrr}\hline
  Variable & \multicolumn{13}{c}{Coef.}\\[2pt]
  \hline
& 1 & 2 & 3 & 4 & 5 & 7 & 8 & 11 & 12 & 13 & 14 & 15 & 17\\[2pt]
  \hline
Christian$^\dagger$ & & & 1.75 & & & & -3.85 & & & & & & \\
Comm. Coll.$^\dagger$ & 2.75 & & & & & & & & & & & & \\
Common App &  & & -1.95 & & & & 2.38 & & 1.75 & & & & \\
Enrollment & & & -0.53 & 1.99 & -0.50 & & -0.73 & & & -0.56 & & & \\
HBCU$^\dagger$ & & & & & & 7.29 & & & & & & & \\
HSI$^\dagger$ &  & & & & & & & 1.30 & & & 1.65 & & \\
Large Online$^\dagger$ & & & & & & & & &  & & & & 3.20\\
Liberal Arts & & & & & & & -1.77 & & & & & 2.97 & \\
No App Fee & 1.19& & & & & & & & & & & & \\
Private$^\dagger$ & & & & & & & -2.88 & & & & & & \\
SAT/ACT Opt. & 1.44 & & -0.74 & & & & & & & & & & -2.08\\
Some Online$^\dagger$ & & & & & 0.92 & & & & -1.61 & & & & \\
Trade School$^\dagger$ & 3.34 & -2.42 & & & & & & & & & & & \\
Tuition & -1.85 & & & & 1.68 & & 3.14 & & -1.05 & & & & \\
Verified & -1.34 & & -1.27 & 1.80 & & & & & -1.27 & & & & 
  \\[2pt]
\hline
\end{tabular}
\end{center}
\end{table}

\begin{figure}[htp]
\centering
\includegraphics[width=\textwidth]{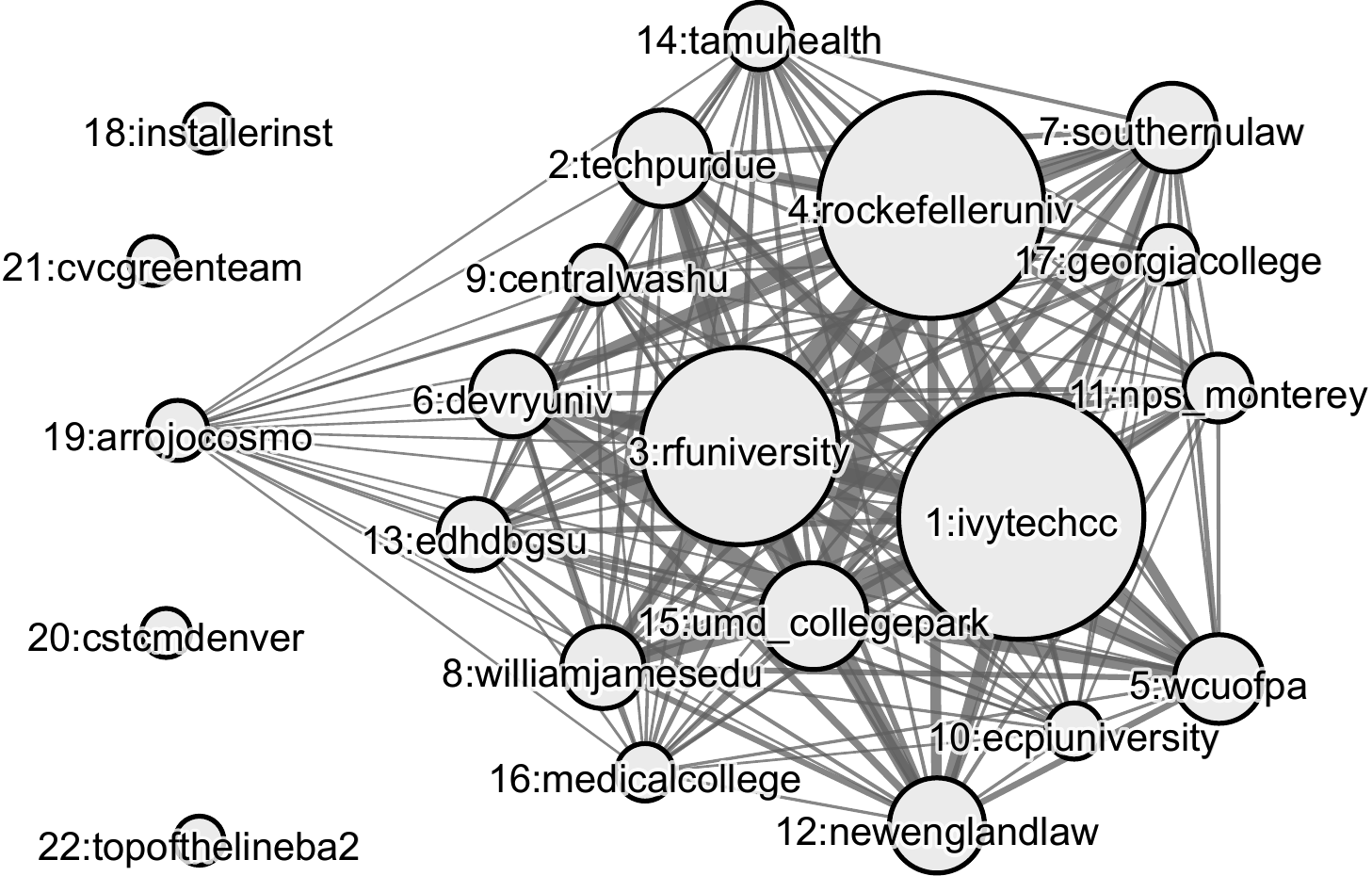}
\caption{\label{figure:induced}An induced network of IHE
  clusters. Each node represents a cluster named after its
  highest-enrollment IHE. The node size represents the number of IHEs
  in the cluster. The edge width represents the number of IHE-level
  connections.}
\end{figure}

As a side note, community detection can be used to visualize
$G$. Large networks are usually hard to visualize, especially when
their Newmann modularity is low, and the community structure is not
prominent. We use the extracted partition $C$ to build a bird's-eye
view of $G$, known as an induced network $I=\left(C,E^I\right)$
(Fig.~\ref{figure:induced}). An induced node in $I$ represents a
cluster in $G$. An induced edge between two nodes $c_i$ and $c_j$ in
$I$ exists iff there exists at least one edge from any node in $c_i$
to any node in $c_j$:

\begin{equation}
  \forall i,j: \exists e^I_{ij} \Leftrightarrow \left(\exists k,l:
  n_k\in c_i\wedge n_l\in c_j\wedge \exists e_{kl}\right).
\end{equation}

Respectively, the weight of such induced edge $w^I_{ij}$ is the number
of the original edges in $G$ from any node in $c_i$ to any node in
$c_j$:

\begin{equation}
  w^I_{ij} = \#\{e_{kl}\,|\,n_k\in c_i\wedge n_l\in c_j\}.
\end{equation}

The name of each cluster in Fig.~\ref{figure:induced} incorporates the
name of the Twitter account of the IHE with the highest enrollment in
the cluster.

\section{\label{sec:glue}Followers' Analysis}
At the last stage of the network analysis, we shift the focus of
attention from the IHEs to their followers.

We selected 14,750 top followers who follow at least 1\% of the IHEs
in our data set. Approximately 8\% of them have an empty description
or a description in a language other than English. Another 268
accounts belong to the IHEs from the original data set, and at least
326 more accounts belong to other IHEs, both domestic and
international.

We constructed a semantic network of lemmatized tokens by connecting
the tokens that frequently (10 or more times) occur together in the
descriptions. We applied the Louvain~\cite{blondel08} community
detection algorithm to extract topics---the clusters of words that are
frequently used together. The algorithm identified twelve topics named
after the first nine most frequently used words. For each follower's
account, we selected the most closely matching topics. The names and
counts for the most prominent topics are shown in
Table~\ref{table:topics}.

\begin{table}
\caption{\label{table:topics}The most prominent topics and the number
  of followers accounts that use them. (Since a description may
  contain words from more than one topic, the sum of the counts is
  larger than the number of followers.) $^\dagger$Topic \#8 is
  technical.}
\begin{center}
\begin{tabular}{llr}\hline
ID&Top seven topic terms & Count\\[2pt]
  \hline
1&education, service, higher, business, professional, solution, research&5,039\\
2&student, program, online, academic, year, helping, opportunity   &  3,460\\
3&school, high, official, twitter, news, account, follow           &  3,308\\
4&college, community, campus, institution, mission, member, black  &  2,870\\
5&help, life, world, love, social, work, people                    &  2,258\\
6&university, career, state, new, job, find, best                  &  1,814\\
7&coach, teacher, author, husband, father, writer, book            &  1,125\\
8$^\dagger$&endorsement, like, link, facebook, retweets, equal, following & 557\\
9&lover, mom, wife, mother, dog                                    &    515\\
[2pt]
\hline
\end{tabular}
\end{center}
\end{table}

Even after the manual cleanup, some of the 12,984 remaining followers'
accounts probably still belong to IHEs and associated divisions,
organizations, and officials. This deficiency would explain the
significance of the topics \#4 and, partially, \#2 that seem to use
the endogenous terminology. The remaining topics are exogenous to the
IHEs and represent higher education services, high schools,
communities, career services, and individuals (``male'' and
``female'').

\section{\label{sec:discussion}Discussion}
Based on the results from Section~\ref{sec:analysis}, we look at each
independent variable's influence on each network and Twitter
performance parameter, whenever the influence is statistically
significant ($p\le0.01$).

It has been observed~\cite{valente2008} that the centrality measures are
often positively correlated. Indeed, in $G$'s case, we saw strong
($\ge0.97$) correlations between the degree, eigenvector, and
closeness centralities, which explains their statistically significant
connection to the same independent variables
(Table~\ref{table:monadic}). More central nodes tend to represent:
\begin{description}
\item[Some specialty IHEs:] Liberal arts colleges, HBCUs.
\item[Internet-savvy IHEs:] IHEs with a longer presence on Twitter,
  IHEs with some or many online programs.
\item[Bigger IHEs with simplified application options:] IHEs with no
  application fees (and accepting Common App---for the closeness
  centrality), larger IHEs.
\end{description}

All these IHEs blend better in their possibly non-homogeneous network
neighborhoods.

The betweenness centrality---the propensity to act as a shared
reference point---is positively affected by being a liberal arts
college or private IHE, and longer presence on Twitter, and negatively
affected by higher tuition and being a Christian IHE. On the contrary,
large and Christian IHEs tend to have a larger local clustering
coefficient and a more homogeneous network neighborhood.

All Twitter performance measures: the numbers of favorites, followers,
friends, and posts---are positively affected by enrollment and the
verified account status. The number of posts is also higher for the
IHEs with a more prolonged presence on Twitter and Christian IHEs. The
number of followers is also higher for private IHEs and lower for
liberal arts colleges and IHEs with some online programs. The latter
observation is counterintuitive and needs further exploration.

Edge weight is the only dyadic variable in
$G$. Table~\ref{table:edges} shows that the weight of an edge is
explained by the differences of the adjacent nodes' attributes. Some
of the attributes promote homophily, while others inhibit it.

The strongest edges connect the IHEs located in the same state, which
is probably because many local IHEs admit the bulk of the local high
schools' graduates and are followed by them and their parents. Much
weaker, but still positive, contributors to the edge weight are
similar enrollment and tuition, same religious affiliation, online
teaching preferences, racial preferences, and application preferences,
a ``classical'' list of characteristics that breed
connections~\cite{mcpherson01}. We hypothesize that prospective
students and their parents follow several IHEs that match the same
socio-economic profile. National, regional, and professional
associations (such as the National Association for Equal Opportunity
and National Association of Independent Colleges and Universities) may
follow similar IHEs for the same reason.

We identified two factors that have a detrimental effect on edge
weight: having the same gender designation (``All-Male,''
``All-Female,'' or neither) and especially the same ``Liberal Arts''
designation. There are 1.58\% of ``All-Female'' IHEs (and no
``All-Male'') and 11.2\% Liberal Arts colleges in our data set. The
IHEs of both types may be considered unique and not substitutable,
thus having fewer shared followers.

In the same spirit, some network communities (clusters) of $G$
represent compact groups of IHEs with unique characteristics
(Table~\ref{table:cluster}). For example, cluster 1 tends to include
community colleges and trade schools with no application fees,
optional SAT/ACT, and lower tuition (e.g., Carl Sandburg
College). Cluster 3 is a preferred locus of smaller Christian IHEs
that do not accept Common App but require SAT/ACT (New Saint Andrews
College). The last comprehensive example is cluster 8: smaller public,
secular, expensive IHEs embracing Common App (University of Maine at
Machias). IHEs with large online programs are in cluster 17 (Middle
Georgia State University), Historically Black Colleges and
Universities---in cluster 7 (North Carolina A\&T State University),
and Liberal Arts colleges---in cluster 15 (St. Olaf College).

It is worth reiterating that the membership in five clusters
containing 9.1\% IHEs, cannot be statistically significantly explained
by any independent variable. The explanatory variables, if they exist,
must be missing from our data set.

\section{\label{sec:conclusion}Conclusion}
We constructed and analyzed a social network of select North American
institutions of higher education (IHEs) on Twitter, using the numbers
of shared followers as a measure of connectivity. We used multiple OLS
regression to explain the network characteristics: centralities,
clustering coefficients, and cluster membership. The regression
variables include IHE size, tuition, geographic location, type, and
application preferences. We discovered statistically significant
connections between the independent variables and the network
characteristics. In particular, we observed strong homophily among the
IHEs in terms of the number of shared followers. Finally, we analyzed
the self-provided descriptions of the followers and assigned them to
several classes. Our findings may help understand the college
application decision-making process from the points of view of the
major stakeholders: applicants, their families, high schools, and
marketing and recruitment companies.

%
%
\bibliographystyle{splncs03}
\bibliography{cs}
\end{document}